# Liutex Core Lines as a New Framework for Tropical Cyclone Vortex Analysis

Yifei Yu, Chenxi Ma, Chaoqun Liu*

Department of Mathematics, University of Texas at Arlington, Arlington, TX, USA

*:cliu@uta.edu

Abstract:

Liutex core lines are systematically applied to Hurricane Dorian to investigate the three-dimensional vortical organization of an intense tropical cyclone. Unlike conventional analysis based on wind speed or vorticity, Liutex core lines identify the local vortex axes and provide a compact geometrical representation of rotational structures. The results show that Dorian is not characterized by a single vortex core, but by a large population of vortices embedded within a coherent hurricane-scale circulation. Stronger and more vertically organized vortices are concentrated in the inner-core region, whereas numerous weaker, inclined, and curved vortical structures extend outward and collectively retain the spiral organization of the storm. The complete core line field therefore reveals a three-dimensional, multiscale vortical skeleton of the hurricane. These results suggest that a tropical cyclone can be viewed as an organized multiscale vortex system rather than a single monolithic vortex. Beyond visualization, Liutex core lines provide a potential framework for quantifying vortex geometry, spatial organization, and temporal evolution, offering a new approach for studying structural changes and vortex dynamics in tropical cyclones.

## 1. Introduction

Tropical cyclones are highly organized atmospheric systems in which rotation, convection, vertical motion, and multiscale flow interactions coexist over a wide range of spatial scales. Their most recognizable structures, including the eye, eyewall, and spiral rainbands, have traditionally been identified from satellite imagery, radar reflectivity and wind fields. These descriptions have been remarkably useful for operational meteorology and hurricane dynamics. Nevertheless, they are primarily based on observable storm features rather than on a rigorous identification of the underlying three-dimensional vortical structures. In particular, the eyewall and rainbands are readily recognizable meteorological features, but their relationship to individual vortices and to the three-dimensional organization of rotational motion remains less clearly investigated.

The difficulty becomes increasingly important as the spatial resolution of numerical simulations and observations improves. At coarse resolution, a tropical cyclone can appear as a relatively simple cyclonic circulation with a broad maximum of vertical vorticity surrounding the storm center. At finer resolutions, however, the flow within the inner core and rainband regions becomes more complex. Localized regions of strong rotation, intense shear, convective updrafts, downdrafts, and filamentary structures coexist within the larger storm circulation. The resulting flow field is therefore better viewed as a multiscale dynamical system than as a single coherent vortex.

Vorticity has long served as one of the principal diagnostics of rotational motion in tropical cyclone studies. Although indispensable in atmospheric dynamics, vorticity is not a direct measure of local fluid rotation because it contains contributions from both rotation and shear [1–3]. This distinction is particularly relevant in hurricanes, where strong velocity gradients are ubiquitous in the eyewall, boundary layer, rainbands, and regions associated with convective structures. Large vorticity may consequently occur in strongly sheared regions without identifying the center or axis of an actual vortex. Conventional vortex criteria [4–6] developed in fluid mechanics can provide additional information, but most are scalar quantities and are commonly visualized through contours or isosurfaces. Such representations indicate where a criterion exceeds a selected level, yet they do not directly provide the geometrical axis, connectivity, or spatial trajectory of individual vortices.

Liutex [7,8] was introduced as a physical quantity for identifying the local rotational component of fluid motion and separating it from shear [9]. Rather than treating vorticity as synonymous with rotation, Liutex distinguishes rigid rotational motion from the non-rotational contributions contained in the velocity gradient. This distinction provides a more direct description of local vortex strength and direction. The application of Liutex to tropical cyclones by Alvarez et al. [10] demonstrated that hurricane circulation contains numerous vortical structures distributed across different regions and spatial scales. Their results suggested that a tropical cyclone should not be interpreted simply as one large atmospheric vortex. Instead, the large-scale circulation contains a hierarchy of smaller vortices embedded within and interacting with the parent storm.

A remaining challenge is how to describe the geometry and organization of these vortices in three dimensions. Scalar vortex quantities can identify rotational regions, but they do not by themselves reveal how individual vortices extend through space, how their axes are oriented, or how vortices in the eyewall and rainbands are arranged relative to the larger hurricane circulation. Isosurfaces can provide three-dimensional visualization, but their geometry depends on the selected threshold and complex structures may merge, split, or obscure one another. For a flow containing a large population of vortices, these limitations make it difficult to extract an interpretable representation of the underlying vortical architecture.

Liutex core lines [11] offer a fundamentally different way of representing such structures. A Liutex core line follows the local rotation axis and therefore provides a one-dimensional representation of the center of a three-dimensional vortex. Instead of displaying the entire volume occupied by rotational motion, the core-line description extracts the geometrical skeleton of the vortex system. The resulting lines provide direct information about vortex location, orientation, spatial extent, and organization while greatly reducing the complexity of the three-dimensional flow field. This representation is particularly attractive for tropical cyclones, in which vortices spanning multiple scales coexist within a very large computational or observational domain.

In the present study, Liutex core lines are systematically applied to the full three-dimensional flow field of Hurricane Dorian to characterize the storm-wide vortical structures. The analysis focuses on how core lines represent the dominant circulation surrounding the hurricane eye, the vortical structures embedded within the eyewall, and the numerous smaller structures extending into the outer circulation and spiral rainband regions. Their spatial organization is compared with conventional flow diagnostics and observed storm morphology to determine what additional structural information can be obtained from a vortex-axis representation.

The purpose of this work is therefore broader than introducing another visualization technique. By converting a complicated three-dimensional rotational field into a collection of physically defined vortex axes, Liutex core lines provide a new framework for examining the internal architecture of tropical cyclones. Such a representation may enable individual vortical structures to be identified, tracked, classified, and quantitatively compared during storm evolution. It also creates a pathway toward studying how changes in vortex organization are related to processes such as eyewall evolution, rainband development, vortex interaction, structural change, and tropical cyclone intensification. The storm-wide application of Liutex core lines thus provides a new geometrical perspective for investigating tropical cyclones as organized multiscale vortex systems.

## 2. Liutex and Liutex Core Lines

Vorticity is conventionally used to characterize rotational motion in fluid flows. However, vorticity does not represent fluid rotation alone because it contains contributions from both rigid rotation and shear. This distinction becomes particularly important in highly sheared flows such as tropical cyclones, where large velocity gradients occur not only within vortices but also across the eyewall, boundary layer, and convective structures. Liutex was developed to isolate the local rotational component of the flow from these non-rotational contributions.

When a three-dimensional velocity gradient tensor $\nabla \boldsymbol{u}$ has one real eigenvalue and a pair of complex-conjugate eigenvalues, the normalized real eigenvector $\boldsymbol{r}$ defines the local rotation axis. Its direction is chosen such that

$$\boldsymbol{\omega} \cdot \boldsymbol{r} > 0 \tag{1}$$

where $\boldsymbol{\omega}$ is the vorticity vector.

The Liutex magnitude can be calculated by [12]

$$R = \boldsymbol{\omega} \cdot \boldsymbol{r} - \sqrt{(\boldsymbol{\omega} \cdot \boldsymbol{r})^2 - 4\lambda_{ci}^2} \tag{2}$$

where $\lambda_{ci}$ is the imaginary part of the complex-conjugate eigenvalues of $\nabla \boldsymbol{u}$. The Liutex vector is then

$$\boldsymbol{R} = R\boldsymbol{r} \tag{3}$$

$\boldsymbol{r}$ gives the local axis of rotation, while $R$ is twice angular speed. In contrast to vorticity, Liutex removes the contribution associated with shear and consequently provides a more direct measure of local fluid rotation.

Liutex nevertheless remains a local quantity. A three-dimensional Liutex field provides the strength and direction of local rotation at every point but does not, by itself, describe how these locally rotating regions are connected to form individual vortices. For a complex flow containing a large number of vortical structures, an additional geometrical representation is needed to extract the center and spatial trajectory of each vortex.

A three-dimensional vortex possesses a local rotation axis. The Liutex core line [11] is constructed to represent the geometrical skeleton of a three-dimensional vortical structure. At the center of a vortex, the Liutex magnitude reaches a local extremum in the plane normal to the rotation axis. The variation of $R$ in directions perpendicular to the rotation axis therefore vanishes. Consequently, the $\nabla R$ must be parallel to the local rotation axis,

$$\nabla R/\quad/\boldsymbol{r} \tag{4}$$

Equivalently, the Liutex core line condition can be written as

$$\nabla R \times \boldsymbol{r} = 0 \; and \; R > 0 \tag{5}$$

Liutex core lines are lines formed by points satisfying Eq. (5).

The Liutex core line representation differs fundamentally from conventional vortex visualization based on contours or isosurfaces. An isosurface describes a finite region corresponding to a prescribed value of a scalar vortex identification method, and its appearance may change substantially with the selected threshold. A core line instead represents the center of the rotational structure. Once extracted, it directly describes the position, orientation, curvature, and spatial extent of the vortex without displaying the entire vortex volume.

This reduction from a three-dimensional vortical region to a one-dimensional core line is particularly useful for tropical cyclone analysis. A hurricane contains vortical structures over a broad range of spatial scales, and simultaneous visualization of these structures using three-dimensional isosurfaces can become obscured by overlapping volumes. Liutex core lines substantially reduce this geometrical complexity. Large and small vortices can be represented simultaneously by their rotational axes, allowing their spatial distribution and organization within the eye, eyewall, rainbands, and outer circulation to be examined directly.

For an isolated vortex, a single Liutex core line provides a compact description of its rotational center. For a tropical cyclone containing a large population of vortices, the collection of Liutex core lines provides a representation of the vortical skeleton of the storm. This representation forms the basis for the tropical cyclone analyses presented in the following sections.

## 3. Liutex Core-Line Structure of Hurricane Dorian

Hurricane Dorian is used to demonstrate how Liutex core lines can reveal the internal vortical organization of an intense tropical cyclone. The analysis is based on the storm at 13:00 UTC on 1 September 2019. The resulting core-line field provides a compact representation of the locations, orientations, and strengths of vortical structures within the hurricane.

3.1 Top view of the Vortex System

Figure 1(a) presents a top view of the Liutex core lines around the center of Hurricane Dorian. A clear cyclonic organization is immediately visible. The core lines are not randomly distributed throughout the domain. Instead, they collectively form a spiral pattern centered on the hurricane eye, revealing the organization of numerous individual vortical structures within the large-scale storm circulation.

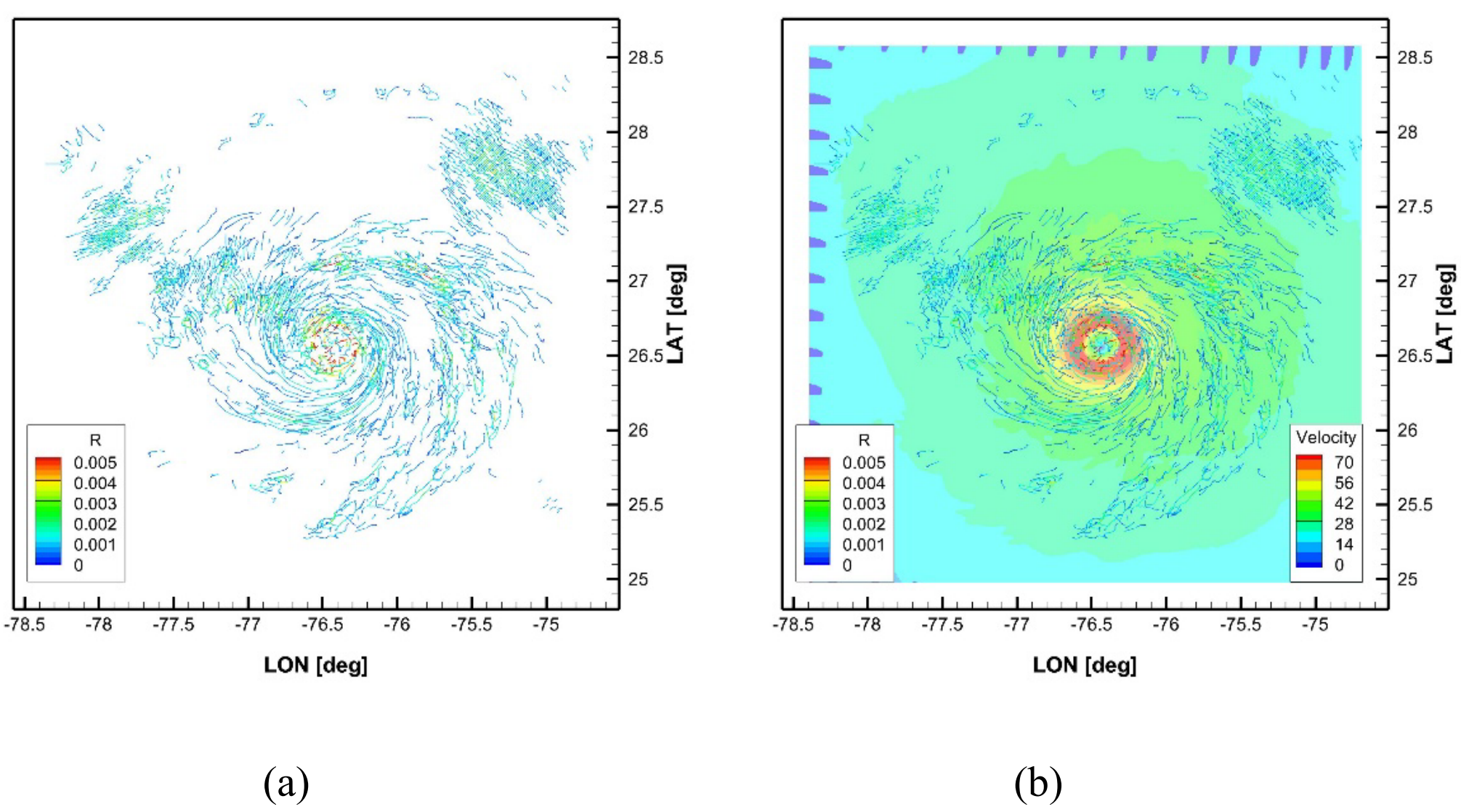

Fig. 1 Top view of Hurricane Dorian at 13:00 UTC on 1 September 2019. (a) Liutex core lines. (b) Liutex core lines and 900 hPa wind speed distribution.

The strongest core lines, indicated by relatively large Liutex magnitude, are concentrated in the inner-core region. Moving outward from the storm center, the Liutex magnitude generally decreases, while a much larger population of weaker core lines extends into the surrounding circulation. Many of these outer structures follow curved and spiral trajectories around the hurricane center. The large-scale morphology of the storm is therefore retained even though the rotational field has been reduced to a collection of individual vortex axes.

An important feature of Fig. 1(a) is the absence of a single core line representing the entire hurricane. If Dorian were interpreted literally as one isolated vortex, one might expect a dominant vortex axis passing approximately through the storm center. The Liutex core lines instead reveal a large number of distinct rotational structures. The hurricane-scale circulation emerges from their collective spatial organization rather than from a single vortex core.

Figure 1(b) superimposes the Liutex core lines on the horizontal wind-speed field at 900 hPa. The wind-speed distribution shows the familiar tropical cyclone structure, including a relatively low-speed eye surrounded by a region of substantially stronger winds. The strongest Liutex core lines are concentrated primarily within and around this inner-core high-wind region.

3.2 Three-Dimensional Vortical Structure

The top view reveals the overall organization of the vortex system but cannot describe the orientation and vertical development of individual vortices. Figure 2(a) therefore presents the complete three-dimensional Liutex core line field.

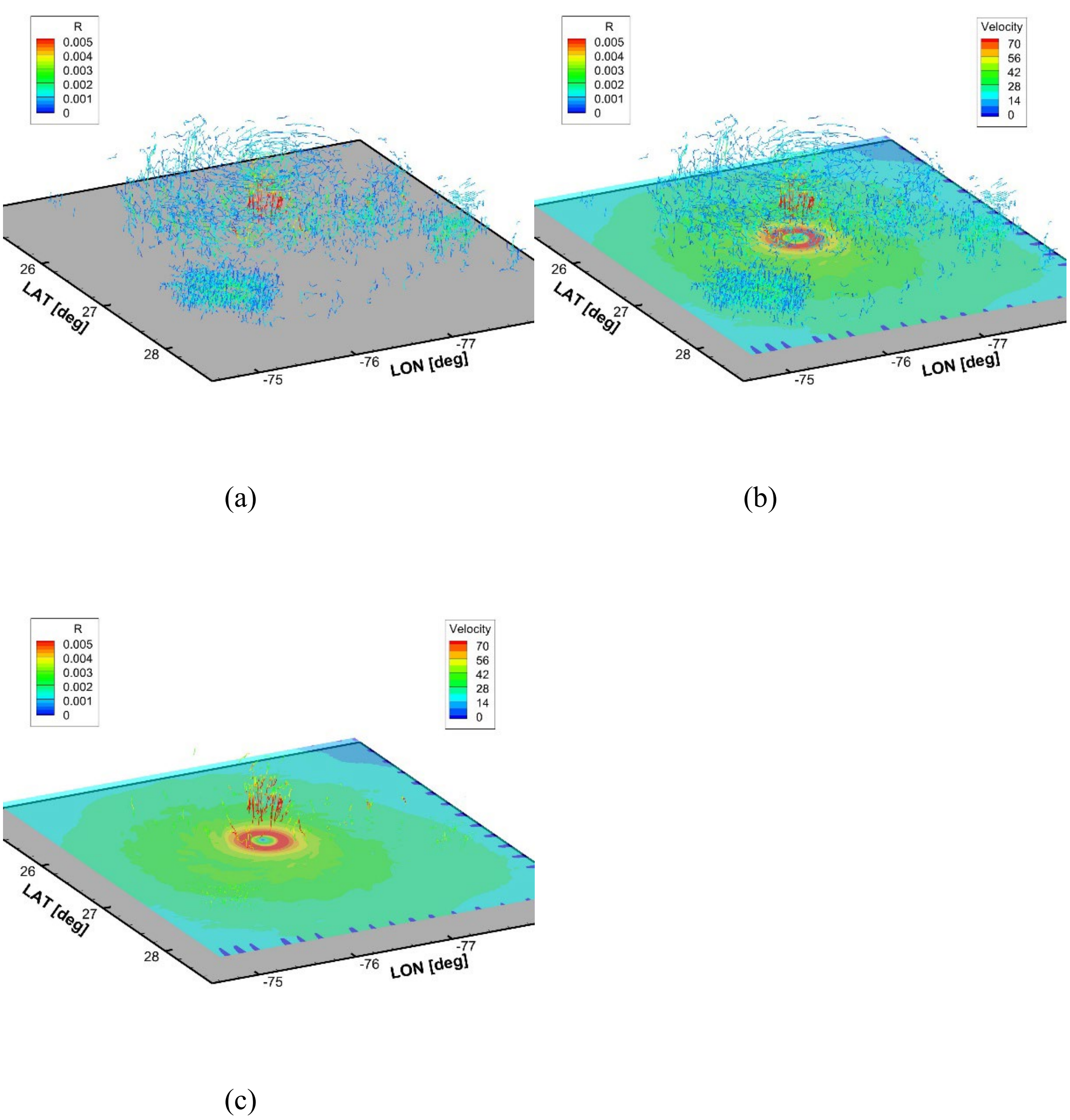

Fig. 2 Three-dimensional organization of Hurricane Dorian at 13:00 UTC on 1 September 2019. (a) Liutex core lines. (b) Liutex core lines and 900 hPa wind velocity distribution. (c) Dominant vertical vortices and 900 hPa wind velocity distribution.

The resulting structure is considerably more complex than would be inferred from a two-dimensional representation. The core lines exhibit a wide range of orientations and geometries, including nearly vertical, inclined, curved, and approximately horizontal structures. This demonstrates that the hurricane vortex system is intrinsically three-dimensional.

A pronounced concentration of strong core lines appears in the inner-core region. Many of these structures possess a substantial vertical component and extend through multiple levels of the analyzed flow field. Their Liutex magnitudes are also considerably larger than those of most vortices farther from the storm center. The inner core therefore contains not only the strongest winds but also some of the strongest and most vertically organized local rotational structures.

Outside the inner core, the core line geometry becomes increasingly heterogeneous. Numerous weaker vortices are distributed throughout the surrounding circulation. Their axes may be tilted, curved, or organized into clusters with similar orientations. Some groups of core lines also retain the spiral organization evident in the horizontal projection. These structures demonstrate that the outer circulation is not simply a smoothly rotating background flow but contains a large population of distinct three-dimensional vortices.

Figure 2(b) combines the three-dimensional core-line field with the 900-hPa wind-speed distribution. This representation illustrates the different information contained in the two diagnostics. The wind field emphasizes the hurricane-scale circulation and the radial variation of wind speed, whereas the core lines reveal the discrete vortical structures distributed within and above that circulation.

The combined visualization also shows that the large-scale cyclonic organization persists across different spatial scales. At the storm scale, the complete population of core lines is organized around the hurricane center. At smaller scales, each line describes the axis of an individual local vortex. Liutex core lines therefore provide a direct geometrical connection between local vortical structures and the hurricane-scale circulation.

### 3.3 Dominant Inner-Core Vortices and the Multiscale Vortical Skeleton

Because the complete core line field contains a very large number of vortical structures, the strongest inner-core features can be partly obscured by the surrounding weaker vortices. Figure 2(c) highlights the dominant core lines with comparatively strong Liutex magnitude and a pronounced vertical orientation.

Once the weaker structures are suppressed, a compact group of strong vortex axes becomes clearly visible near the hurricane center. A particularly important observation is that the inner-core rotational structure is not represented by a single dominant vertical core line. Instead, multiple strong core lines coexist in the central region of the storm. Their axes are predominantly vertical but exhibit variations in position, orientation, and curvature.

This result provides a different geometrical interpretation of the hurricane inner core from the idealized picture of a single vortex rotating about one central axis. The instantaneous flow contains multiple local vortices embedded within the larger cyclonic circulation. The hurricane can therefore possess one dominant large-scale circulation without being a single vortex in the local kinematic sense.

The distinction between these two descriptions is important. A tropical cyclone clearly exhibits coherent rotation at the storm scale, but coherent large-scale circulation does not require the entire flow to share one vortex axis. Liutex core lines show that the large-scale storm circulation contains a hierarchy of vortical structures with different strengths, sizes, orientations, and spatial locations.

The inner and outer regions also display different characteristics. Stronger and more vertically organized vortices are concentrated near the storm center, whereas weaker and geometrically more diverse vortices extend outward into the surrounding circulation. Despite these differences, the individual structures are not independent of the hurricane-scale organization. Their collective arrangement retains the cyclonic and spiral geometry of the parent storm.

The complete Liutex core-line field can therefore be interpreted as a three-dimensional vortical skeleton of Hurricane Dorian. The skeleton preserves the large-scale organization of the hurricane while simultaneously resolving the individual vortices from which that organization is constructed. This representation reveals Dorian as an organized multiscale vortex system rather than as a single monolithic vortex.

More generally, the core line representation provides quantities that can be analyzed directly rather than inferred from scalar contours or isosurfaces. Individual vortices can, in principle, be characterized by their location, length, orientation, curvature, and Liutex strength, and their evolution can be followed from one time instant to another. This creates a basis for examining how the organization of the vortex system changes during tropical cyclone evolution and how such structural changes may be related to changes in storm intensity and morphology.

## 4. Conclusion

Liutex core lines were applied to Hurricane Dorian to examine the three-dimensional organization of vortical structures within an intense tropical cyclone. The method reduces the complex rotational field to a set of vortex axes, providing direct information on their location, orientation, strength, and spatial organization.

The results show that Dorian is not represented by a single vortex core. Instead, the hurricane contains numerous vortical structures embedded within one coherent cyclonic circulation. Stronger and more vertically organized vortices are concentrated in the inner-core region, while weaker and more diverse structures extend outward and retain the spiral organization of the storm.

These results suggest that a tropical cyclone is better viewed as an organized multiscale vortex system rather than a single monolithic vortex. The complete Liutex core-line field can therefore be interpreted as the three-dimensional vortical skeleton of the hurricane.

Beyond visualization, Liutex core lines also provide a potential framework for quantifying vortex strength, orientation, length, curvature, spatial distribution, and temporal evolution. This may offer a new approach for studying structural changes and vortex dynamics during tropical cyclone evolution.